\documentclass[prd,twocolumn]{revtex4-1}
\usepackage{amsmath}
\usepackage{amssymb}
\usepackage{graphicx}
\usepackage{bm}
\usepackage{enumerate}
\usepackage{color}
\usepackage[colorlinks = true,
            linkcolor = blue,
            urlcolor  = blue,
            citecolor = blue,
            anchorcolor = blue]{hyperref}
\usepackage{physics}

\newcommand{\be}{\begin{equation}}
\newcommand{\ee}{\end{equation}}
\newcommand{\bea}{\begin{eqnarray}}
\newcommand{\eea}{\end{eqnarray}}
\newcommand{\ba}{\begin{eqnarray}}
\newcommand{\ea}{\end{eqnarray}}

\begin{document}

\title{Are there monopoles in the quark-gluon plasma?}
\author{Adith Ramamurti}
\email[]{adith.ramamurti@stonybrook.edu}
\author{Edward Shuryak}
\email[]{edward.shuryak@stonybrook.edu}
\author{Ismail Zahed}
\email[]{ismail.zahed@stonybrook.edu}
\affiliation{ Department of Physics and Astronomy, \\ Stony Brook University,\\
Stony Brook, NY 11794, USA}

\date{\today}

\begin{abstract} 
Monopole-like objects have been identified in multiple lattice studies, and there is now  a significant amount of literature on their importance in phenomenology. Some analytic indications of their role, however, are still missing. The 't Hooft-Polyakov monopoles, originally derived in the Georgi-Glashow model, are an important dynamical ingredient in theories with extended supersymmetry ${\cal N} = 2,\,4$, and help explain the issues related with electric-magnetic duality. There is no such solution in QCD-like theories without scalar fields.  However, {\em all} of these theories have instantons and their finite-$T$ constituents known as instanton-dyons (or instanton-monopoles). The latter leads to semiclassical partition functions, which for  ${\cal N} = 2,\,4$ theories were shown to be identical (``Poisson dual") to the partition function for monopoles.  We show how, in a pure gauge theory, the semiclassical instanton-based partition function can also be Poisson-transformed
into a partition function, interpreted as the one of moving and rotating monopoles.
 \end{abstract}

\maketitle
\section{Introduction}
The possible existence of magnetic monopoles in electrodynamics fascinated leading physicists in the 19th century. With the development of quantum mechanics, Dirac \cite{Dirac:1931kp} related the existence of monopoles with the electric charge quantization.  However, QED monopoles were never found.
  
Classical solitons with magnetic charge were found by 't Hooft \cite{NUPHA.B79.276} and Polyakov \cite{Polyakov:1974ek} in the Georgi-Glashow model. Such monopoles exist and play an important role in other theories with an adjoint scalar field, notably in theories with extended supersymmetry ${\cal N} = 2,\,4$. 
Their presence and properties have significantly advanced our understanding of the electric-magnetic duality and its relation to the renormalization group (RG) flow. In the  ${\cal N} = 2$ case, there is a gradual transition from an electric theory at weak coupling to a magnetic theory at strong coupling \cite{Seiberg:1994rs}. In the  ${\cal N} = 4$ case, monopoles dressed by bound fermions  were shown to create an ${\cal N} = 4$ multiplet of fields, making the electric and magnetic  theories the same, up to a coupling. This implies that the beta function of $g$ and $1/g$ must be the same, therefore just zero, explaining why this theory must be conformal. 

In QCD-like theories without scalars, e.g. pure gauge theories or ${\cal N} = 1$ SYM, there are no such monopole solutions. 
Despite this,  Nambu \cite{Nambu:1974zg}, 't Hooft \cite{NUPHA.B190.455}, and Mandelstam \cite{Mandelstam:1974pi} proposed the ``dual superconductor" model of the electric color confinement. In this model, the Bose-Einstein condensation (BEC) of monopoles at $T\leq T_c$ expels electric fields from the vacuum into confining flux tubes. 

In lattice studies of gauge theories, monopoles have been identified, and their locations and paths were correlated with gauge-invariant observables, such as the action and square of the magnetic field \cite{Laursen:1987eb}.  The monopoles were found to create a magnetic current around the electric flux tube \cite{Koma:2003gq, Bornyakov:2003vx}. In Landau gauge, while monopole-type singularities themselves are not present, the physical properties that they source are still present and gauge-invariant \cite{Suzuki:2009xy}. The motion and correlations of the monopoles were shown to be as expected for a Coulomb plasma \cite{DAlessandro:2007lae, Bonati:2013bga, Liao:2008jg}, the deconfinement critical temperature $T_c$  does coincide accurately with that of monopole BEC transition \cite{D'Alessandro:2010xg,Bonati:2013bga,Ramamurti:2017fdn}, and the BEC transition has been shown to be gauge independent \cite{Bonati:2010tz, Bonati:2010bb, DiGiacomo:2017blx}.
 
The ``magnetic scenario" for quark-gluon plasma (QGP) \cite{Liao:2006ry,Liao:2007mj,Liao:2008jg} assumes the presence of non-condensed monopoles as another kind of quasiparticles. Unlike quarks and gluons, which have vanishing densities at  $T\rightarrow T_c$, the monopole density has a peak near $T_c$. Monopole-gluon and monopole-quark scattering were shown to play a significant role in kinetic properties of the QGP, such as the shear viscosity $\eta$ \cite{Ratti:2008jz} and the jet quenching parameter $\hat q$  \cite{Xu:2015bbz,Xu:2014tda,Ramamurti:2017zjn}. The non-condensed monopoles should also lead to electric flux tubes at  $T>T_c$  \cite{Liao:2007mj}, which were recently observed on the lattice \cite{Cea:2017ocq}. Thus, there is a growing amount of phenomenological evidence suggesting magnetic monopoles do exist, not only as a confining condensate at $T\leq T_c$, but also as non-condensed quasiparticles at $T>T_c$. While the central role of monopoles in the confinement-deconfinement transition was recognized long ago, their relation to another important non-perturbative aspect of QCD-like theories, chiral symmetry breaking, has attracted much less attention prior to our recent paper \cite{Ramamurti:2018hdh}, in which we have demonstrated how the quark condensate is formed as $T\rightarrow T_c$.

Nevertheless, this phenomenological evidence does not convince many theorists, who would rather have an analytic argument not relying on lattice numerics or heavy-ion phenomenology. One such argument will be provided by this paper. It is still {\em indirect}, in the sense that we do not have a microscopic description of these monopoles in terms of the gauge fields. We do, however, derive the corresponding partition function, based on a transformed  semiclassical partition function.
%

The semiclassical description of the vacuum of gauge theories is based on the instanton solution \cite{Belavin:1975fg}. At finite temperatures, however, the 4d instanton solution has been shown to dissolve into instanton constituents, known as  instanton-dyons (or instanton-monopoles) \cite{Kraan:1998kp,Kraan:1998pm,Lee:1998bb}.  Studies of the ensembles of instanton-dyons have explained the deconfinement and chiral symmetry restoration transitions both numerically \cite{Larsen:2015vaa,Larsen:2015tso} and using a mean-field analysis \cite{Liu:2015ufa,Liu:2015jsa}. For a recent short review, see Ref. \cite{Shuryak:2017kct}.

The construction of the instanton-dyons starts from the same 't Hooft-Polyakov monopole, but with the zeroth component of the gauge field $A_0$ acting as the scalar adjoint ``Higgs" field.  However, these objects are {\em pseudo}-particles and not particles, existing only in the Euclidean formulation of the theory for which $A_0$ is real. Therefore, while  instanton-dyons do lead to successful semiclassical applications, their usage for phenomenological applications is severely limited. Another obstacle to their development, perhaps even more important for many, is that their physical meaning remains rather obscure. In this paper, we argue that it should not be so, and that the instanton-dyon  gauge field configurations are nothing else but {\em quantum paths of moving and rotating monopoles}. 
 
A gradual understanding of this statement began some time ago, but remained rather unnoticed by the larger community. One reason for that was the setting in which it was shown, which was based on extended supersymmetry. Only in these cases was one able to derive reliably {\em both} partition functions -- in terms of monopoles and instanton-dyons -- and show them to be equal \cite{Dorey:2000qc,Poppitz:2011wy,Poppitz:2012sw}. Furthermore, they were not summed up to an analytic answer, but shown instead to be related by the so-called ``Poisson duality."

Since this concept it also not widely known, Sec. \ref{sec_rotator} contains a pedagogical section, which discusses a much simpler toy model of a rotator -- a particle on a circle -- at finite temperature. We also obtain two expressions for its partition function, 
one based on its excited states and one based on ``winding paths" in Euclidean time. In this model, one can derive the analytic solution for both sums and directly see that they are the same. 

In Sec. \ref{sec_susy}, we turn to theories with extended supersymmetry. This section is a brief pedagogical review of the works of Dorey and collaborators, and shows how  the Poisson duality works in this case -- almost identically to the rotator model. 

In Sec. \ref{sec_pure_gauge}, we turn to pure gauge theories at finite temperature, using as above its simplest version with  SU(2) color. We will explicitly derive the $n$-winding  gauge configurations, periodic on the Matsubara circle, and the corresponding semi-classical partition function. We then Poisson-transform it into another form, the one we argue is counting occupations
of the excited states of moving/rotating monopoles. 

\section{Quantum rotator at finite $T$ and its dual descriptions}\label{sec_rotator}

A quantum rotator is a particle moving on a circle. Its location is defined by the angle $\alpha\in[0,2\pi]$ and its action is defined by kinetic and topological parts 
\be  S=\oint \dd t\, {\Lambda\over 2}\, \dot\alpha^2+ S_\text{top}(\omega)\,, \ee
where $\dot \alpha=\dd\alpha/\dd t$, and $\Lambda=m R^2$ is the corresponding moment of inertia for rotation. (It can be
set to unity by appropriately selecting units, but for the purposes of this paper, we keep it.)

The topological part $S_\text{top}\sim \int \dd t \dot \alpha(t) $ does not lead to any ``force" -- there is no contribution to the classical equation of motion -- but it provides an extra phase factor in the quantum partition function. The phenomenon was introduced by Aharonov and Bohm in a celebrated paper \cite{Aharonov:1959fk} and is well known. We remind the reader that
this phase is an external parameter which can be induced by a solenoid in extra dimensions, provided the rotating particle is charged and the time derivative is generalized to the long gauge-invariant derivative including the $A_0$ field.
 
The  quantum mechanical spectrum of states is immediately obtained via quantization of the angular momentum $l$ and the partition function at temperature $T$ is
\be   Z_1=\sum_{l=-\infty}^\infty \exp \bigg(-{l^2 \over 2\Lambda T}+i l \omega  \bigg)\,,
\ee  
where, for convenience, we normalized the  Aharonov-Bohm contribution to a phase $\omega$. Since the angular momentum $l$ is integer-valued, each term in $Z_1$ is a periodic function of this phase, with a natural $2\pi$ period. Note also that positive and negative $l$ cancel the imaginary part, so $Z_1$ is real. Finally, this sum is best convergent at small temperature $T$, where only a few states close to the ground state with  $l=0$ need to be included.
 
In the dual approach, finite temperature is introduced via the standard Euclidean Matsubara time defined on {\em another} circle $\tau\in [0,\beta\equiv1/T]$. The path integral which leads to the partition function needs to be done over the periodic paths, $\alpha(0)=\alpha(\beta)$, so one may say that the Euclidean theory is a particle on a double torus. 

Classes of paths which make a different number $n$ of rotations around the original circle can be defined as ``straight" classical periodic paths
\be \alpha_n(\tau)=2\pi n {\tau \over \beta}\,, \ee 
plus small fluctuations around them. Carrying out a Gaussian integral over them leads to the following
partition function,
\be  Z_2=\sum_{n=-\infty}^\infty \sqrt{2\pi \Lambda T}\, \exp \bigg( -{T \Lambda \over 2}(2\pi n - \omega)^2 \bigg)\,. 
\ee
The key point here is that these quantum numbers,
$l$ used for $Z_1$ and $n$ for $Z_2$, are very different in nature.
In $Z_1$, each term of the sum is periodic in $\omega$, while in $Z_2$, this property is recovered only after summation over $n$.
 The temperature $T$  in $Z_2$ happens to be in an unusual place, in the numerator of the exponent, so this sum converges best at {\em high} temperature, unlike 
 the sum in $Z_1$. Indeed, at high $T$ the Matsubara circle becomes small and the path integral is dominated by paths with small number of windings. 

\begin{figure}[ht]
\begin{center}
\includegraphics[width=\linewidth]{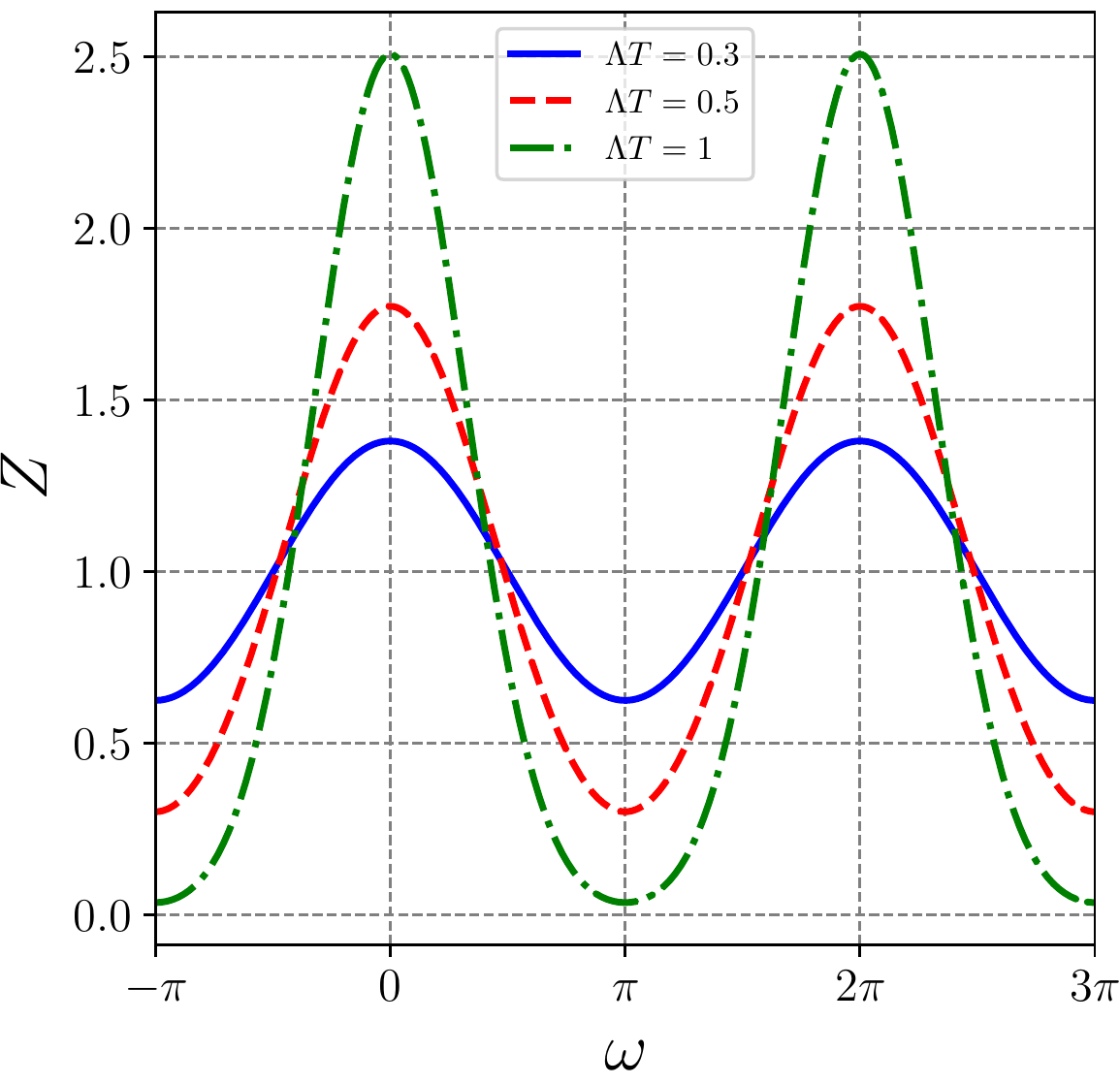}
\caption{The partition function $Z$ of the rotator as a function of the external
Aharonov-Bohm phase $\omega$ (two periods are shown to emphasize its periodicity). 
The (blue) solid, (red) dashed and (green) dash-dotted curves are for $\Lambda T=0.3,0.5,1.$}
\label{fig_rotator_Z}
\end{center}
\end{figure}

In spite of such differences, both expressions are in fact the same! In this toy model, it is not difficult to do the sums numerically and plot the results. Furthermore, one can also derive the analytic expressions, expressible in terms of the elliptic theta function
of the third kind 
\be Z_1 = Z_2=  \theta_3\bigg(-{\omega \over 2},\, \exp\bigg(-{ 1 \over 2\Lambda T}\bigg) \bigg)\,, \ee
which is plotted in Fig. \ref{fig_rotator_Z} for few values of the temperature $T$. 

Mathematically, the identity of the two sums can be traced to the fact that our path integral is defined on two circles, or, equivalently, a 2d torus, and the circles can be interchanged. In string theory, such relations are known as T-duality. In practice, these are the low-temperature and the high-temperature approximations, often used without noticing the exact summation and duality.

Even if one is not able to identify the sums as the same elliptic function, the equality can be seen from the observation that the sum $Z_1$ is the discrete Fourier transform of the Gaussian, which is known to be the ``periodic Gaussian" appearing in $Z_2$.  One can further recognize that the identity of the two sums is just a particular case of a more general relation known in mathematics as the {\em Poisson summation formula}, valid not only for a Gaussian but for arbitrary functions. For reference, let us mention here one particular version \cite{Zygmund},
\be \sum_{n=-\infty}^\infty f(\omega+n P) =\sum_{l=-\infty}^\infty {1 \over P} \tilde f \bigg({l\over P}\bigg)e^{i 2 \pi l \omega/P}\,, \label{eqn_Zygmund}
\ee 
where $f(x)$ is some function, $\tilde f$ is its Fourier transform, and $P$ is the period of both sums
as a function of the ``phase" $\omega$.

\section{Semiclassical theory and monopoles in theories with extended supersymmetry}
 \label{sec_susy}
\subsection{The setting}

All of the following discussion concerns a Euclidean theory defined on $R^3\times S^1$. In this section, unlike in the following one, all of the fields, including the fermions, have periodic boundary conditions on $S^1$, and therefore supersymmetry is {\em not} broken.

We study the weak coupling $g\ll 1$ scenario, which makes the instantons and their constituents -- as well as the monopoles with actions/masses $O(1/g^2)$ -- heavy enough to trust the dilute gas approximation. This lets us focus on a single object and avoid finite-density (many-body) complications. In the ${\cal N}=4$ theory, the charge does not run and $g$ is simply an input parameter. In the  ${\cal N}=2$ theory, however, the coupling does run, and one needs to select the circumference of the circle $\beta$ to be small enough such that the corresponding frequencies $\sim 2\pi/\beta$ are large enough to ensure weak coupling. 

Compactification of one coordinate to the circle is needed to introduce ``holonomies,'' gauge invariant integrals over the circle
$ \oint \dd x_\mu A^\mu,\,  \oint \dd x_\mu C^\mu$ of the electric and magnetic potentials, respectively. Their values can have nonzero expectation values, which can be viewed as external parameters given by Aharonov-Bohm fluxes through the circle induced by fields in extra dimensions. These holonomies will play important role in what follows. Dorey et al. \cite{Dorey:2000dt} call these external parameters $\omega$ and $\sigma$, respectively.

Finally, in order to make the discussion simpler, one assumes the minimal non-Abelian color group SU($N_c$) with the number of colors $N_c=2$. This group has only one single diagonal generator $\tau^3$, breaking the color group SU(2)$\rightarrow$U(1).

The theories with extended supersymmetry ${\cal N} = 2,\,4$ have one and six adjoint scalar fields, respectively.  Recall that these two theories also have, respectively, 2 and 4 fermions, so that the balance between bosonic and fermionic degrees of freedom is perfect.  For simplicity, all vacuum expectation values (VEV) of the scalars, as well as both holonomies are assumed to be in this diagonal direction, so the scalar VEVs and  $\omega$ and $\sigma$ are single-valued parameters without indices. In the general  group SU($N_c$), the number of diagonal directions is the Abelian subgroup, and thus the number of parameters is $N_c-1$.

\subsection{Monopoles and their partition function}
 
Considering the theories on the Coulomb branches, with nonzero VEVs of the scalars $\phi_A, A=1\ldots6$, one can naturally use the original BPS version of the 't Hooft-Polyakov monopole, with a mass
\be M=\bigg({4 \pi \over g^2}\bigg)\phi\,. \ee 

We will only discuss the ${\cal N}=4$ case, following  Dorey and collaborators \cite{Dorey:2000dt}. Six scalars and two holonomies can be combined to vacua parameterized by 8 scalars, extended by supersymmetry to 8 chiral supermultiplets. These 8 fermions have zero modes, describing their binding to monopoles. We will, however, not discuss any of those in detail.

%
%

%

The SU(2) monopole has four collective coordinates, three of which are related with translational symmetry and location in space, while the fourth is rotation around the $\tau^3$ color direction, 
\be \hat \Omega={\rm exp}(i\alpha \hat \tau^3/2)\,. \ee
 Note that such rotation leaves unchanged the presumed VEVs of the Higgses and holonomies, as well as the Abelian $A^3_\mu\sim 1/r$ tails of the monopole solution. Nevertheless, these rotations are meaningful because they do rotate  the monopole core -- made up of non-Abelian $A^1_\mu,\,A^2_\mu$ fields -- nontrivially. It is this rotation in the angle $\alpha$ that makes the monopole problem similar to a quantum rotator.  As was explained by Julia and Zee \cite{Julia:1975ff}, the corresponding integer angular momentum is nothing but the electric charge of the rotating monopole, denoted by $q$. 

Now that we understand the monopoles and their rotated states, one can  define the partition function at certain temperature, which (anticipating the next sections) we will call $T\equiv1/\beta$,
\bea 
\label{Zmono}
Z_\text{mono}&&= \sum_{k=1}^\infty \sum_{q=-\infty}^\infty\bigg( {\beta \over g^2}\bigg)^8 
{k^{11/2} \over \beta^{3/2} M^{5/2} }  \nonumber\\
&&\times \exp\bigg( i k \sigma - i q \omega -\beta k M -{\beta \phi^2 q^2 \over 2 k M}\bigg)\nonumber\,,\\
\eea
where $k$ is the magnetic charge of the monopole. The derivation can be found in the original paper, and we only comment that the temperature in the exponent only appears twice, in the denominators of the mass and the rotation terms, as expected. The two other terms in the exponent, $\exp( i k \sigma - i q \omega)$, are the only places where holonomies appear, as the phases picked up by magnetic and electric charges over the circle.  

\subsection{Instantons and monopole-dyons} 

Now we derive an alternative 4d version of the theory, in which we will look at gauge field configurations in all coordinates including  the compactified  ``time coordinate" $\tau$. These objects are versions of instantons, split by a nonzero holonomy into instanton constituents.  Since these gauge field configurations need  to be periodic on the circle, and this condition can be satisfied by paths adding arbitrary number $n$ of rotations, their actions are
 \be 
 S_\text{mono}^n=\bigg({4 \pi \over g^2}\bigg)\bigg({\beta^2 |\phi|^2+| \omega - 2\pi n |^2} \bigg)^{\frac 12}\,,
 \ee 
including the contribution from the scalar VEV $\phi$, the electric holonomy $\omega$, and the winding number of the path $n$. In the absence of the holonomies, the first term would be $M/T$ as one would expect. 

The partition function then takes the form \cite{Dorey:2000dt}
\bea
&& Z_\text{inst}= \sum_{k=1}^\infty \sum_{n=-\infty}^\infty\bigg( {\beta \over g^2} \bigg)^9{k^6 \over (\beta M)^3} \nonumber\\
&&\times \exp\bigg( i k \sigma-\beta k M -{k M \over 2 \phi^2 \beta}(\omega - 2\pi n)^2 \bigg)\,,\nonumber\\
\label{Zinst}
\eea
where $M=(4\pi\phi/g^2)$, the BPS monopole mass without holonomies; thus the second term in the exponent is interpreted as just the Boltzmann factor. The ``temperature" appears in the unusual place in the last term (like for the rotator toy model). The actions of the instantons are large at high-$T$ (small circumference $\beta$); the semiclassical instanton theory works best
at high-$T$. 

%
%
%
%

The Poisson duality relation between these two partition functions, Eqs. (\ref{Zmono}) and (\ref{Zinst}), was originally pointed out by Dorey and collaborators \cite{Dorey:2000dt}. In this paper, it was explained earlier for the simpler toy model of a quantum rotator.  Its mathematical origins  were presumably clarified enough by our toy model, and it is perhaps enough to remind the reader that  the two circles (or the double torus) at play are  the angle $\alpha\in [0,2\pi]$ related with the rotation of the monopole in ordinary/color space and the compactified coordinate $\tau\in[0,\beta]$.
 
\section{Semiclassical theory and monopoles in pure gauge theories} \label{sec_pure_gauge}

Now consider theories without adjoint scalars, which do not have an obvious 't Hooft-Polyakov monopole solution. One example of such a theory discussed in Ref. \cite{Dorey:2000dt} is the $\cal N$=1$^*$  theory obtained from the $\cal N$=4 theory by giving a mass to the three chiral multiplets, which, in the IR, eliminates 3 out of 4 fermions and all 6  scalars. We will not discuss this particular case, but proceed directly to pure gauge theory, starting from the instantons.

\subsection{Finite temperature  instanton-dyons with an arbitrary time winding}

At zero temperature, the Euclidean space $R^4$ is symmetric in all four coordinates, and thus the corresponding saddle points of the integral over fields -- the instantons -- are 4d spherically symmetric. At finite temperatures, Euclidean time is defined on the circle $\tau\in [0,\beta]$. The corresponding solitons -- the calorons -- are deformed periodic instantons. 

In order to keep the weak coupling and the small density approximation valid, we need to consider sufficiently high $T$. What this means practically will be discussed at the end of the paper. For simplicity, for now  we will also ignore the issue of a dynamically generated potential and mean value of the electric holonomy on the time circle, and continue to consider it to be an external parameter; we are therefore considering a ``deformed" gauge theory.

%
%

The presence of the holonomy is known to split the calorons into $N_c$ constituents \cite{Kraan:1998kp,Kraan:1998pm,Lee:1998bb} known as instanton-dyons (or instanton-monopoles). The holonomy eigenvalues $\mu_i, i=1\ldots N_c$ enter the gluon and instanton-dyon masses via their differences $\nu_i=\mu_{i+1}-\mu_{i}$. We will consider only the simplest case of the number of colors $N_c=2$, in which case there is a single holonomy parameter. The caloron is composed of two types of the self-dual dyons, known as the time-independent $M$ dyon and the time-twisted $L$ dyon \cite{Diakonov:2004jn}. 

Following the discussion above, we need to consider a larger set of saddle-point configurations with all possible periodic paths. To be explicit, let us derive the corresponding semiclassical configurations. One starts with the static BPS monopole, with the $A_0$ component of the gauge field now as the adjoint scalar. In the simplest ``hedgehog" gauge, the gauge fields are 
\bea
 A^a_4&=&n_a v \bigg({\rm coth}(vr)-{1 \over vr}\bigg) \,, \nonumber\\
 A^a_i&=&\epsilon_{aij} {n_j \over r} \bigg(1- {vr \over {\rm sinh}(vr)}\bigg) \,,
\eea
where $n_a=x_a/r$ is the spatial unit vector and $v$ is the VEV of $A_4$ at large distances
$r\rightarrow \infty$. 

The twisted solution is obtained in two steps. The first is the substitution 
\be v\rightarrow n (2\pi/\beta) -v\,, \ee
and the second is the gauge transformation with the gauge matrix
\be \hat\Omega= {\rm exp}\bigg( - \frac i\beta n \pi  \tau \hat\sigma^3\bigg)\,, \ee
where we recall that $\tau=x^4\in [0,\beta]$ is the Matsubara time. The derivative term in the gauge transformation adds a constant to $A_4$ which cancels out the unwanted $n (2\pi/\beta) $ term,  leaving $v$, the same as for the original static monopole. After ``gauge combing" of $v$ into the same direction, this configuration -- we will call $L_n$ -- can be combined with any other one. The solutions are all self-dual, but the magnetic and (the Euclidean) electric charges are negative for positive $n$, opposite to the original BPS monopole $M$ for which both are positive.

The action corresponding to this solution is \be S_n=(4\pi/g^2)| 2\pi n /\beta -v|\,.\ee The contribution to the partition function 
requires the calculation of the pre-exponent, due to quantum fluctuations around the $L_n$ solution. Following Appendix C of Ref. \cite{Larsen:2015vaa}, this can be extracted from the contribution of the $L$ dyon, which in turn was derived from the explicit calculation of the moduli for the finite temperature instanton ($M$+$L$ system) in Ref. \cite{Diakonov:2004jn}. For the color SU(2) group, taking the limit of large separation the $L$ dyon, the density has the form  
\be 
\dd Z_L \sim \dd^3 x_L \bigg( {8\pi^2 \over g^2} \bigg)^2 e^{-\big( {8\pi^2 \over g^2} \big) \bar \nu }
\big( 2\pi \bar \nu \big)^{8\bar \nu/3} \,,
\ee
with $\bar\nu=1-\nu$ and $\nu=vT/2\pi$. Unlike the theories with extended supersymmetry, there are no cancellations 
in the determinant of the nonzero modes between bosons and fermions, and for $L_n$ classical configurations those have not yet been calculated explicitly. On general grounds, it is expected that it should append the part from the moduli such that the correct running coupling at the relevant scale $\sim 2\pi T \bar \nu$ is reproduced. This means that one expects the exponent to read 
\be 
\dd Z_L \sim \dd^3 x_L  \exp\bigg(-  \bar \nu {8\pi^2 \over g_0^2} +\bar \nu  {22\over 3} {\rm log}\bigg({p_0 \over 2\pi T \bar \nu}\bigg)\bigg) \,,
\ee
where the coupling $g_0$ is the defined at the normalization scale $p_0$. Similarly, the power of the action in numerator must be appended by the two-loop corrections to the two-loop beta function, and so on. 

 For our subsequent discussion, we will ignore the running and only keep the first term, taking the mean coupling to be just a constant at a characteristic $p_0=2\pi T \langle \bar \nu \rangle$, say  
\be S_0\equiv S_L+S_M={8\pi^2 \over g_0^2}=10\,. \ee
The simulation of instanton-dyon ensembles \cite{Larsen:2015vaa} were done for $S_0$ ranging from 5 to 13, and thus defining a rather large range of dyon densities. Higher-twist instantons $L_n$ for  $n>1$ or $n<0$ are all strongly suppressed and in practice can be ignored; the instanton-dyon ensemble calculations performed in Ref. \cite{Larsen:2015vaa} only included the $n=0$ time independent dyon $M$ and the first twisted dyons $L_1$ because, in this range of temperatures, the holonomy phase $\omega$ changes from a small value to $\pi$ at the confining phase transition, where $\omega$ and $2\pi - \omega$ are comparable.

In the present calculation, we will keep all of them, preserving exact periodicity, and write the semiclassical partition function as
\be 
Z_\text{inst}  =\sum_n e^{-\left(\frac{4\pi}{g_0^2}\right) | 2\pi n  - \omega |}
\label{ZINST}
\ee
It is periodic in the holonomy, as it should be. Note that, unlike in Eq. (\ref{Zinst}), it has a modulus rather than a square of the corresponding expression in the exponent. This is due to the fact that the sizes of $L_n$ and their masses are all defined by the same combination $ | 2\pi n -\omega |T$ and therefore the moment of inertia $\Lambda \sim 1/| 2\pi n \beta  - v |$.

\subsection{The Poisson transformation}

A key point of this paper is that the existence of the semiclassical instanton partition function {\em implies the existence of monopoles} moving and rotating in their collective coordinates. According to the general Poisson relation, Eq. (\ref{eqn_Zygmund}),  the Fourier transform of the corresponding function appearing in the sum in Eq. (\ref{ZINST}) reads
\bea
F\left(e^{-A|x|}\right)\equiv&& \int_{\nu=-\infty}^{\infty} \dd x \, e^{i 2\pi\nu -A|x|} \nonumber\\
= &&{2A \over A^2+(2\pi \nu)^2} \,,
 \eea
and therefore the monopole partition function is 
\be 
Z_\text{mono}\sim \sum_{q=-\infty}^\infty e^{i q \omega - S(q) }\,,
\ee
where 
\bea 
S(q)&&={\rm log}\bigg(\bigg({4\pi \over g_0^2}\bigg)^2 + q^2\bigg)\nonumber\\
 &&\approx  2{\rm log}\bigg({4\pi \over g_0^2}\bigg)+q^2 \bigg({g_0^2\over 4\pi}\bigg)^2+\ldots \,,
 \label{eqn_action_rotatingx} 
\eea
where the last equality is for $q \ll 4\pi/g_0^2$.

\section{What have we learned about QCD monopoles?} 

Before summarizing our answer to this question, let us first recall the setting and conclusions of the preceding section. The coupling is presumed small, so $4\pi/g_0^2 \gg 1$ and the semiclassical calculation is well controlled. This implies that the corresponding temperature is ``high enough.'' The holonomies $\omega,\,\sigma$, treated as external Aharonov-Bohm phases imposed on the system, create a certain ``Higgsing" of the gluons, with only the diagonal ones remaining massless. Calorons are split into the instanton-dyons, and the semiclassical partition function, appended by all $L_n$ contributions, can be calculated. 
 
What we would actually like to study is QCD with quarks at temperatures around the deconfinement transition $T\sim T_c$. Indeed, heavy-ion collisions create matter with $T$ between roughly $2T_c\approx 300$ MeV and $0.5T_c$. Most finite-$T$ lattice studies are devoted to this temperature range as well. While the coupling seems to be small enough to keep the semiclassical approach reasonable, $S_0=8\pi^2/g^2\sim 10$, when including the pre-exponent, one finds that the ensemble is not really dilute, and 
in order to perform the integration over the collective variables, one needs to solve a nontrivial many-body problem of a dense instanton-dyon plasma. The instanton-dyon ensemble in this scenario does shift the potential for the electric holonomy dynamically to its ``confining" value, for $T<T_c$. Semiclassical ensembles of instanton-dyons also explain chiral symmetry breaking, and their changes with flavor-dependent quark periodicity phases. Further development of the semiclassical theory is, therefore, well justified.
 
The main point of this paper, however, is different: {\em any} semiclassical partition function, once derived, can be Poisson-rewritten into an identical form, with the sum over certain physical states. We have shown how one can do so for pure gauge theory, without scalars, using a relatively simple, or even schematic, form of its semiclassical partition function, for which we calculated its Poisson dual. We further argued that the resulting partition function can be interpreted as being generated by {\em moving and rotating monopoles}. 

The results are a bit surprising. First, the action of a monopole, although still formally large in weak coupling, is only a logarithm of the semiclassical parameter; these monopoles are therefore quite light. Second is the issue of monopole rotation. The very presence of an object that admits rotational states implies that the monopole core is not spherically symmetric.
The Poisson-rewritten partition function has demonstrated that the rotating monopoles are {\em not} the rigid rotators, because their action, Eq. (\ref{eqn_action_rotatingx}), depends on the angular momentum $q$ and is quadratic only for small values of $q$.
The slow (logarithmic) increase of the action with $q$ implies that the dyons are in fact shrinking with increased rotation. In the moment of inertia, this shrinkage is more important than the growth in the mass, as the size appears quadratically. As strange as it sounds, it reflects on the corresponding behavior of the instanton-dyons $L_n$ with the increasing $n$. 

Although such rotations are well known in principle as Julia-Zee dyons with {\em real} electric charge (unlike that of the instanton-dyons, which only exist in the Euclidean world) and studied in theories with extended supersymmetries, to our knowledge the existence of multiple rotational states of monopoles has not yet been explored in monopole-based phenomenology. In particular, one may wonder how the existence of multiple rotational states affects their Bose-condensation at $T<T_c$, the basic mechanism behind the deconfinement transition. The electric charges of the rotating monopoles  should, therefore, also contribute to the jet quenching parameter $\hat q$ and the viscosity, which was not included before.



%
%
%

We note that perhaps a useful ``middle ground'' connecting the two regimes -- the idealized semiclassical dilute gas and the real-life finite-$T$ QCD -- would be lattice studies of the  gauge theory with an {\em induced holonomy} at high $T$. To our knowledge, this has not been done in detail.  

The semiclassical studies in this direction by introduction of certain masses, by Dorey et al. using ${\cal N}=4  \rightarrow {\cal N}=1$ \cite{Dorey:2000dt} and by Unsal et al. for ${\cal N}=2$ deformed toward pure gauge theory \cite{Poppitz:2011wy}, were the first steps in this direction. Yet the conjectured continuity of both the confined and deconfined phases, from dilute to dense regimes, were never studied nor confirmed. 

Lattice measurements of the holonomy potential for the SU(2) and SU(3) gauge theories with fixed external holonomy have been performed in Ref.\cite{Diakonov:2012dx}. The perturbative renormalized potential was derived and compared to these data in Ref. \cite{Diakonov:2013lja}. Such a subtraction opens the door to studies of the monopole contribution, which has not been attempted as of yet. Other lattice studies, of the ``deformed QCD" with an extra holonomy-dependent term in the action, were performed in Ref. \cite{Ogilvie:2014nra}. 

Finally, the introduction of light quarks allows study of zero modes of both semiclassical instanton-dyons and monopoles, giving another way to test their mutual correspondence.



\begin{thebibliography}{99}


\bibitem{Dirac:1931kp} 
  P.~A.~M.~Dirac,
  Proc.\ Roy.\ Soc.\ Lond.\ A {\bf 133}, 60 (1931).
  doi:10.1098/rspa.1931.0130


\bibitem{NUPHA.B79.276} 
  G.~'t Hooft,
  Nucl.\ Phys.\ B {\bf 79}, 276 (1974).
  doi:10.1016/0550-3213(74)90486-6


\bibitem{Polyakov:1974ek} 
  A.~M.~Polyakov,
  JETP Lett.\  {\bf 20}, 194 (1974)
  [Pisma Zh.\ Eksp.\ Teor.\ Fiz.\  {\bf 20}, 430 (1974)].


\bibitem{Seiberg:1994rs} 
  N.~Seiberg and E.~Witten,
  Nucl.\ Phys.\ B {\bf 426}, 19 (1994)
  Erratum: [Nucl.\ Phys.\ B {\bf 430}, 485 (1994)]
  doi:10.1016/0550-3213(94)90124-4, 10.1016/0550-3213(94)00449-8
  [hep-th/9407087].


\bibitem{Nambu:1974zg} 
  Y.~Nambu,
  Phys.\ Rev.\ D {\bf 10}, 4262 (1974).
  doi:10.1103/PhysRevD.10.4262


\bibitem{NUPHA.B190.455} 
  G.~'t Hooft,
  Nucl.\ Phys.\ B {\bf 190}, 455 (1981).
  doi:10.1016/0550-3213(81)90442-9


\bibitem{Mandelstam:1974pi} 
  S.~Mandelstam,
  Phys.\ Rept.\  {\bf 23}, 245 (1976).
  doi:10.1016/0370-1573(76)90043-0


\bibitem{Laursen:1987eb} 
  M.~L.~Laursen and G.~Schierholz,
  Z.\ Phys.\ C {\bf 38}, 501 (1988).
  doi:10.1007/BF01584402


\bibitem{Koma:2003gq} 
  Y.~Koma, M.~Koma, E.~M.~Ilgenfritz, T.~Suzuki and M.~I.~Polikarpov,
  Phys.\ Rev.\ D {\bf 68}, 094018 (2003)
  doi:10.1103/PhysRevD.68.094018
  [hep-lat/0302006].


\bibitem{Bornyakov:2003vx} 
  V.~G.~Bornyakov {\it et al.} [DIK Collaboration],
  Phys.\ Rev.\ D {\bf 70}, 074511 (2004)
  doi:10.1103/PhysRevD.70.074511
  [hep-lat/0310011].


\bibitem{Suzuki:2009xy} 
  T.~Suzuki, M.~Hasegawa, K.~Ishiguro, Y.~Koma and T.~Sekido,
  Phys.\ Rev.\ D {\bf 80}, 054504 (2009)
  doi:10.1103/PhysRevD.80.054504
  [arXiv:0907.0583 [hep-lat]].


\bibitem{DAlessandro:2007lae} 
  A.~D'Alessandro and M.~D'Elia,
  Nucl.\ Phys.\ B {\bf 799}, 241 (2008)
  doi:10.1016/j.nuclphysb.2008.03.002
  [arXiv:0711.1266 [hep-lat]].


\bibitem{Bonati:2013bga} 
  C.~Bonati and M.~D'Elia,
  Nucl.\ Phys.\ B {\bf 877}, 233 (2013)
  doi:10.1016/j.nuclphysb.2013.10.004
  [arXiv:1308.0302 [hep-lat]].


\bibitem{Liao:2008jg} 
  J.~Liao and E.~Shuryak,
  Phys.\ Rev.\ Lett.\  {\bf 101}, 162302 (2008)
  doi:10.1103/PhysRevLett.101.162302
  [arXiv:0804.0255 [hep-ph]].


\bibitem{D'Alessandro:2010xg} 
  A.~D'Alessandro, M.~D'Elia and E.~V.~Shuryak,
  Phys.\ Rev.\ D {\bf 81}, 094501 (2010)
  doi:10.1103/PhysRevD.81.094501
  [arXiv:1002.4161 [hep-lat]].


\bibitem{Ramamurti:2017fdn} 
  A.~Ramamurti and E.~Shuryak,
  Phys.\ Rev.\ D {\bf 95}, no. 7, 076019 (2017)
  doi:10.1103/PhysRevD.95.076019
  [arXiv:1702.07723 [hep-ph]].


\bibitem{Bonati:2010tz} 
  C.~Bonati, A.~Di Giacomo, L.~Lepori and F.~Pucci,
  Phys.\ Rev.\ D {\bf 81}, 085022 (2010)
  doi:10.1103/PhysRevD.81.085022
  [arXiv:1002.3874 [hep-lat]].


\bibitem{Bonati:2010bb} 
  C.~Bonati, A.~Di Giacomo and M.~D'Elia,
  Phys.\ Rev.\ D {\bf 82}, 094509 (2010)
  doi:10.1103/PhysRevD.82.094509
  [arXiv:1009.2425 [hep-lat]].


\bibitem{DiGiacomo:2017blx} 
  A.~Di Giacomo,
  arXiv:1707.07896 [hep-lat].


\bibitem{Liao:2006ry} 
  J.~Liao and E.~Shuryak,
  Phys.\ Rev.\ C {\bf 75}, 054907 (2007)
  doi:10.1103/PhysRevC.75.054907
  [hep-ph/0611131].


\bibitem{Liao:2007mj} 
  J.~Liao and E.~Shuryak,
  Phys.\ Rev.\ C {\bf 77}, 064905 (2008)
  doi:10.1103/PhysRevC.77.064905
  [arXiv:0706.4465 [hep-ph]].


\bibitem{Ratti:2008jz} 
  C.~Ratti and E.~Shuryak,
  Phys.\ Rev.\ D {\bf 80}, 034004 (2009)
  doi:10.1103/PhysRevD.80.034004
  [arXiv:0811.4174 [hep-ph]].


\bibitem{Xu:2015bbz} 
  J.~Xu, J.~Liao and M.~Gyulassy,
  JHEP {\bf 1602}, 169 (2016)
  doi:10.1007/JHEP02(2016)169
  [arXiv:1508.00552 [hep-ph]].


\bibitem{Xu:2014tda} 
  J.~Xu, J.~Liao and M.~Gyulassy,
  Chin.\ Phys.\ Lett.\  {\bf 32}, no. 9, 092501 (2015)
  doi:10.1088/0256-307X/32/9/092501
  [arXiv:1411.3673 [hep-ph]].


\bibitem{Ramamurti:2017zjn} 
  A.~Ramamurti and E.~Shuryak,
  Phys.\ Rev.\ D {\bf 97}, no. 1, 016010 (2018)
  doi:10.1103/PhysRevD.97.016010
  [arXiv:1708.04254 [hep-ph]].


\bibitem{Cea:2017ocq} 
  P.~Cea, L.~Cosmai, F.~Cuteri and A.~Papa,
  Phys.\ Rev.\ D {\bf 95}, no. 11, 114511 (2017)
  doi:10.1103/PhysRevD.95.114511
  [arXiv:1702.06437 [hep-lat]].


\bibitem{Ramamurti:2018hdh} 
  A.~Ramamurti and E.~Shuryak,
  arXiv:1801.06922 [hep-ph].


\bibitem{Belavin:1975fg} 
  A.~A.~Belavin, A.~M.~Polyakov, A.~S.~Schwartz and Y.~S.~Tyupkin,
  Phys.\ Lett.\  {\bf 59B}, 85 (1975).
  doi:10.1016/0370-2693(75)90163-X


\bibitem{Kraan:1998kp} 
  T.~C.~Kraan and P.~van Baal,
  Phys.\ Lett.\ B {\bf 428}, 268 (1998)
  doi:10.1016/S0370-2693(98)00411-0
  [hep-th/9802049].


\bibitem{Kraan:1998pm} 
  T.~C.~Kraan and P.~van Baal,
  Nucl.\ Phys.\ B {\bf 533}, 627 (1998)
  doi:10.1016/S0550-3213(98)00590-2
  [hep-th/9805168].


\bibitem{Lee:1998bb} 
  K.~M.~Lee and C.~h.~Lu,
  Phys.\ Rev.\ D {\bf 58}, 025011 (1998)
  doi:10.1103/PhysRevD.58.025011
  [hep-th/9802108].


\bibitem{Larsen:2015vaa} 
  R.~Larsen and E.~Shuryak,
  Phys.\ Rev.\ D {\bf 92}, no. 9, 094022 (2015)
  doi:10.1103/PhysRevD.92.094022
  [arXiv:1504.03341 [hep-ph]].


\bibitem{Larsen:2015tso} 
  R.~Larsen and E.~Shuryak,
  Phys.\ Rev.\ D {\bf 93}, no. 5, 054029 (2016)
  doi:10.1103/PhysRevD.93.054029
  [arXiv:1511.02237 [hep-ph]].


\bibitem{Liu:2015ufa} 
  Y.~Liu, E.~Shuryak and I.~Zahed,
  Phys.\ Rev.\ D {\bf 92}, no. 8, 085006 (2015)
  doi:10.1103/PhysRevD.92.085006
  [arXiv:1503.03058 [hep-ph]].


\bibitem{Liu:2015jsa} 
  Y.~Liu, E.~Shuryak and I.~Zahed,
  Phys.\ Rev.\ D {\bf 92}, no. 8, 085007 (2015)
  doi:10.1103/PhysRevD.92.085007
  [arXiv:1503.09148 [hep-ph]].


\bibitem{Shuryak:2017kct} 
  E.~Shuryak,
  arXiv:1710.03611 [hep-lat].


\bibitem{Dorey:2000qc} 
  N.~Dorey and A.~Parnachev,
  JHEP {\bf 0108}, 059 (2001)
  doi:10.1088/1126-6708/2001/08/059
  [hep-th/0011202].


\bibitem{Poppitz:2011wy} 
  E.~Poppitz and M.~Unsal,
  JHEP {\bf 1107}, 082 (2011)
  doi:10.1007/JHEP07(2011)082
  [arXiv:1105.3969 [hep-th]].


\bibitem{Poppitz:2012sw} 
  E.~Poppitz, T.~SchŠfer and M.~Unsal,
  JHEP {\bf 1210}, 115 (2012)
  doi:10.1007/JHEP10(2012)115
  [arXiv:1205.0290 [hep-th]].


\bibitem{Aharonov:1959fk} 
  Y.~Aharonov and D.~Bohm,
  Phys.\ Rev.\  {\bf 115}, 485 (1959).
  doi:10.1103/PhysRev.115.485


\bibitem{Zygmund}
A.Zygmund, Trigonometric series, Cambridge University Press, 2nd ed, 1968

\bibitem{Dorey:2000dt} 
  N.~Dorey,
  JHEP {\bf 0104}, 008 (2001)
  doi:10.1088/1126-6708/2001/04/008
  [hep-th/0010115].


\bibitem{Julia:1975ff} 
  B.~Julia and A.~Zee,
  Phys.\ Rev.\ D {\bf 11}, 2227 (1975).
  doi:10.1103/PhysRevD.11.2227


\bibitem{Diakonov:2004jn} 
  D.~Diakonov, N.~Gromov, V.~Petrov and S.~Slizovskiy,
  Phys.\ Rev.\ D {\bf 70}, 036003 (2004)
  doi:10.1103/PhysRevD.70.036003
  [hep-th/0404042].

\bibitem{Diakonov:2012dx} 
  D.~Diakonov, C.~Gattringer and H.~P.~Schadler,
  JHEP {\bf 1208}, 128 (2012)
  doi:10.1007/JHEP08(2012)128
  [arXiv:1205.4768 [hep-lat]].

\bibitem{Diakonov:2013lja} 
  D.~Diakonov, V.~Petrov, H.~P.~Schadler and C.~Gattringer,
  JHEP {\bf 1311}, 207 (2013)
  doi:10.1007/JHEP11(2013)207
  [arXiv:1308.2328 [hep-lat]].

\bibitem{Ogilvie:2014nra} 
  M.~Ogilvie and P.~Meisinger,
  PoS LATTICE {\bf 2014}, 339 (2014)
  [arXiv:1411.5344 [hep-lat]].


\end{thebibliography}
\end{document}